\documentclass[a4paper,11pt]{article}
\usepackage{pos}
\usepackage{subcaption}
\usepackage{amsmath}
\usepackage{gensymb} 

\usepackage[mathlines]{lineno} 

\usepackage{xspace}
\usepackage{slashed}

\definecolor{nicered}{rgb}{0.5,0.,0.}
\definecolor{nicegreen}{rgb}{0.,0.5,0.}
\definecolor{niceblue}{rgb}{0.,0.,0.5}

\title{Precision test of the muon-Higgs coupling at a high-energy muon collider}

\author*[1]{J\"urgen Reuter}
\author[2]{Tao Han}
\author[3]{Wolfgang Kilian}
\author[3]{Nils Kreher}
\author[4]{Yang Ma}
\author[3]{Tobias Striegl}
\author[2]{Keping Xie}

\affiliation[1]{Deutsches Elektronen-Synchrotron DESY,
Notkestr. 85, 22607 Hamburg, Germany}

\affiliation[2]{Pittsburgh Particle Physics, Astrophysics, and Cosmology Center, Department of Physics and Astronomy, University of Pittsburgh, Pittsburgh, PA 15206, USA}

\affiliation[3]{Department of Physics, University of Siegen, Walter-Flex-Straße 3, 57068 Siegen, Germany}

\affiliation[4]{INFN, Sezione di Bologna, Via Irnerio 46, 40126 Bologna, Italy}

\emailAdd{juergen.reuter@desy.de}
\emailAdd{than@pitt.edu}
\emailAdd{kilian@physik.uni-siegen.de}
\emailAdd{nils.kreher@uni-siegen.de}
\emailAdd{yang.ma@bo.infn.it}
\emailAdd{tobias.striegl@physik.uni-siegen.de}
\emailAdd{xiekeping@pitt.edu}

\abstract{We explore the sensitivity of directly testing the muon-Higgs coupling at a high-energy muon collider. This is strongly motivated if there exists new physics that is not aligned with the Standard Model Yukawa interactions which are responsible for the fermion mass generation. We illustrate a few such examples for physics beyond the Standard Model. With the accidentally small value of the muon Yukawa coupling and its subtle role in the high-energy production of multiple (vector and Higgs) bosons, we show that it is possible to measure the muon-Higgs coupling to an accuracy of ten percent for a 10 TeV muon collider and a few percent for a 30 TeV machine by utilizing the three boson production, potentially sensitive to a new physics scale about $\Lambda \sim$ 30--100 TeV.}

\FullConference{%
  41st International Conference on High Energy Physics - ICHEP2022\\
  6-13 July, 2022\\
  Bologna, Italy
}

\begin{document}
\renewcommand{\hookAfterAbstract}{%
\par\bigskip
\textsc{DESY 22-187, PITT-PACC 2214, SI-HEP-2022-34}
}
\maketitle

\section{Introduction}

To scrutinize the Higgs boson discovered a decade ago at the Large Hadron Collider (LHC) is of the utmost importance, especially such
elusive parameters like the muon-Higgs Yukawa coupling in order to check whether the Standard Model (SM) Higgs mechanisms is
responsible for the masses of the second lepton generation. There is already first evidence for the existence of a muon-Higgs coupling
from the LHC measurements~\cite{Sirunyan:2020two,Aad:2020xfq}, and the high-luminosity LHC (HL-LHC) will provide such a test with a
precision of maybe even below ten percent, albeit in a model-dependent way from the total rate of production times decay. A future high-energy muon collider operating in the multi-TeV regime (between 3 and up to 30 TeV) can perform a high-precision Higgs program~\cite{Han:2020pif}, and can directly access the muon-Higgs coupling as both
Higgs and longitudinal electroweak (EW) gauge bosons couple through it to the initial state muons. There is a subtle cancellation
between the gauge and Goldstone parts of the amplitudes mandated by the perturbative unitarity of the SM or any of its extensions. Muon
colliders have recently gained, triggered by experimental advances, a lot of interest in the US Snowmass Community
Study~\cite{MuonCollider:2022glg,MuonCollider:2022xlm,MuonCollider:2022nsa,Aime:2022flm,Black:2022cth}.
Specific new physics models like large extra dimensions, Randall-Sundrum models or also Higgs sector or supersymmetric extensions
predict large deviations of the muon-Higgs coupling. Within this study based on~\cite{Han:2021lnp} we investigate a model-independent
approach based on different effective-field theory (EFT) frameworks. The setup will be discussed in the next section.

\section{Model setup and phenomenology}

We parameterized deviations of the muon-Higgs coupling in two different frameworks based
on a linear Higgs realization (SMEFT with truncation at dimensions 6, 8 or 10) and on a
non-linear Higgs realization (HEFT, also truncated). Besides an extreme benchmark where the higher-dimensional contributions exactly cancel the SM coupling resulting in a vanishing muon-Higgs coupling 
("matched" case), we consider deviations of the muon-Higgs coupling parameterized in terms of a signal-strength modifier, $\kappa_\mu$ (for details cf.~\cite{Han:2021lnp}), modelled in an EFT framework with dimension-6 operators only or dimension-6 and dimension-8 operators simultaneously. For more details, cf.~\cite{Han:2021lnp}.

\begin{figure}[t]
\centering
         \includegraphics[width=0.5\textwidth]{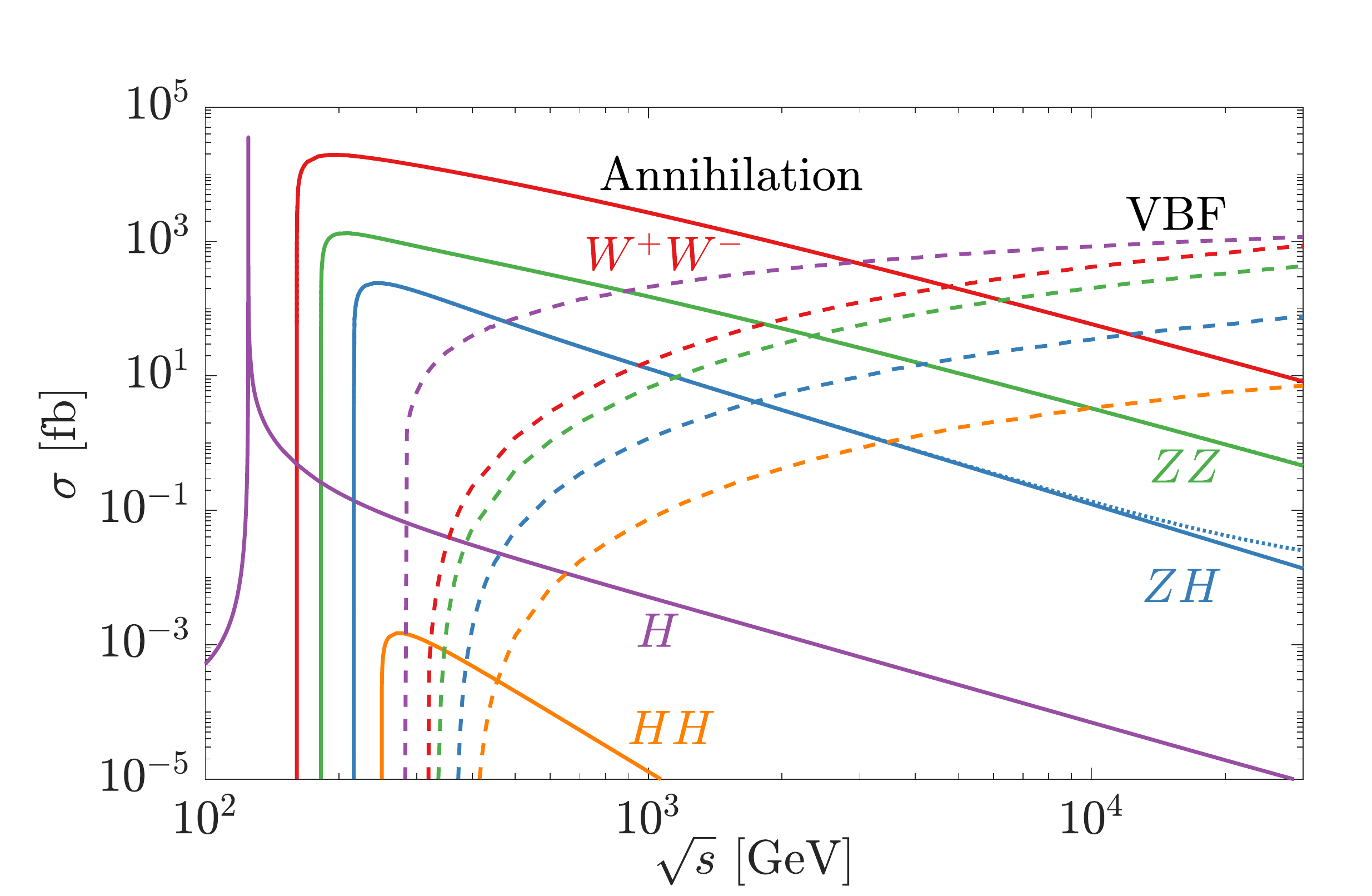}
         \includegraphics[width=0.48\textwidth]{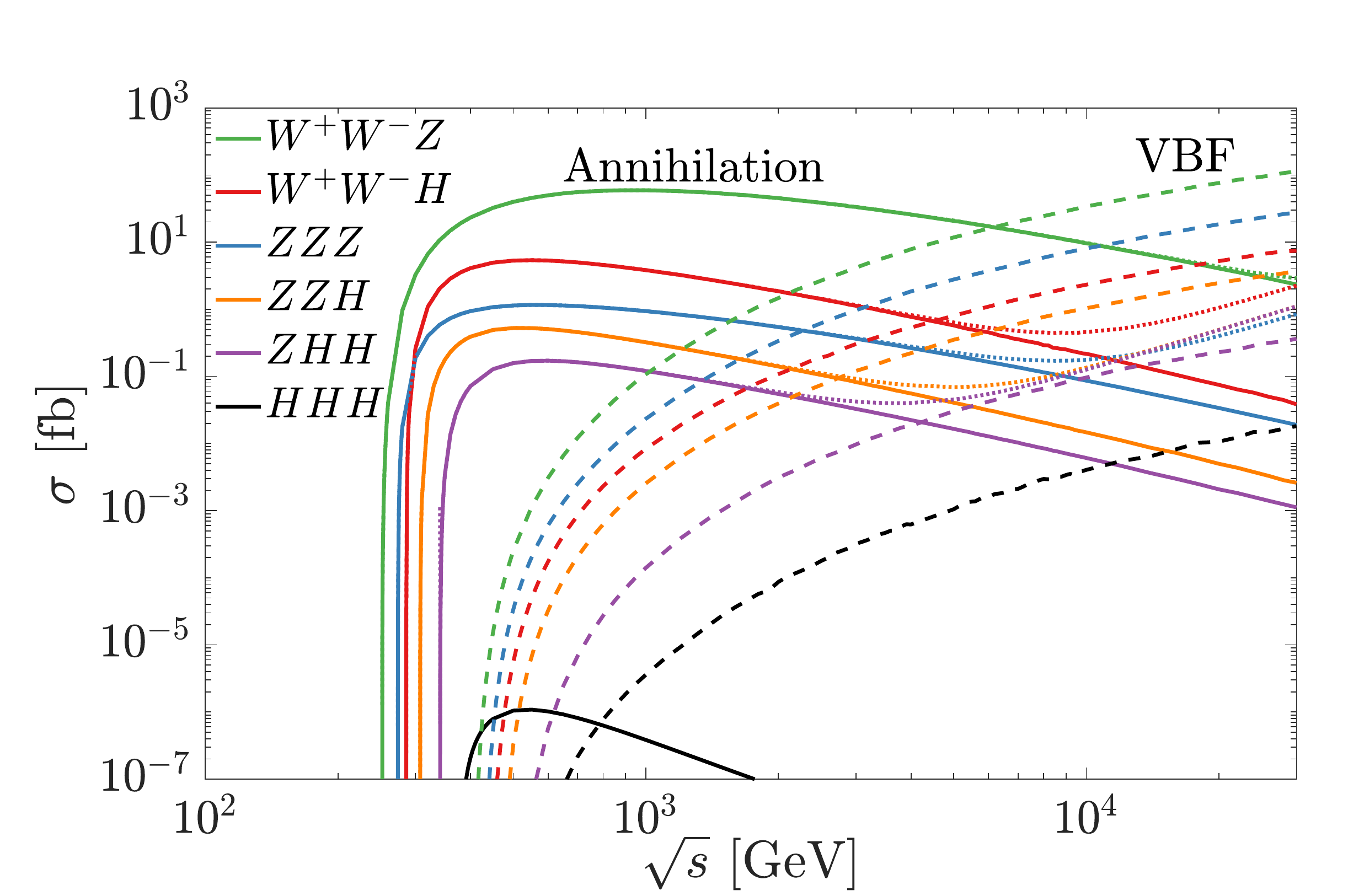}
         \caption{Cross sections for di- and triboson processes on the left and right, respectively, as a function of the muon collider energy. Full lines show prompt production, dashed lines production through vector-boson fusion and dotted lines deviations induced by higher-dimensional operators.}
         \label{fig:xsec}
\end{figure}

\begin{figure}[t]
\centering
         \includegraphics[width=0.48\textwidth]{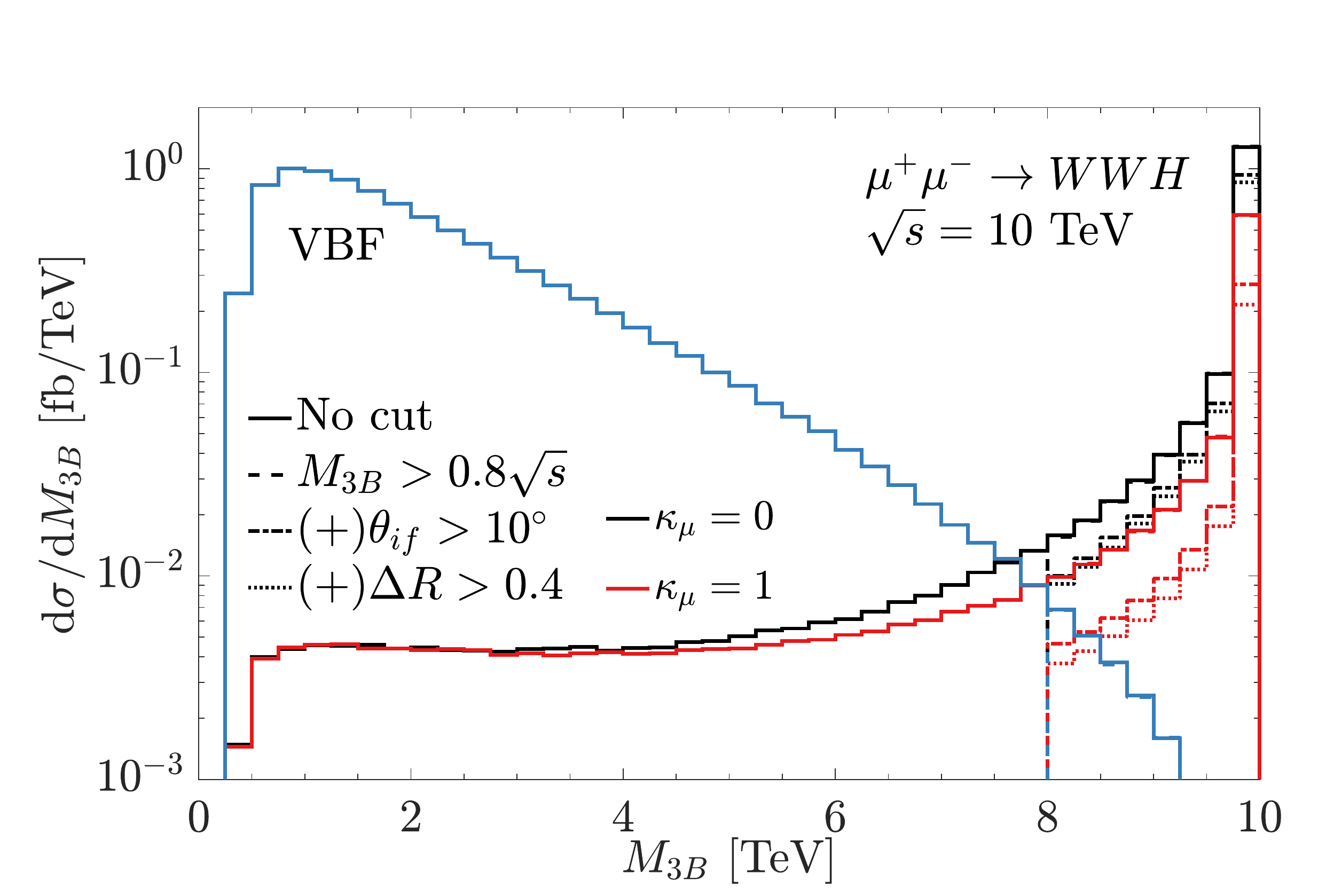}
         \includegraphics[width=0.48\textwidth]{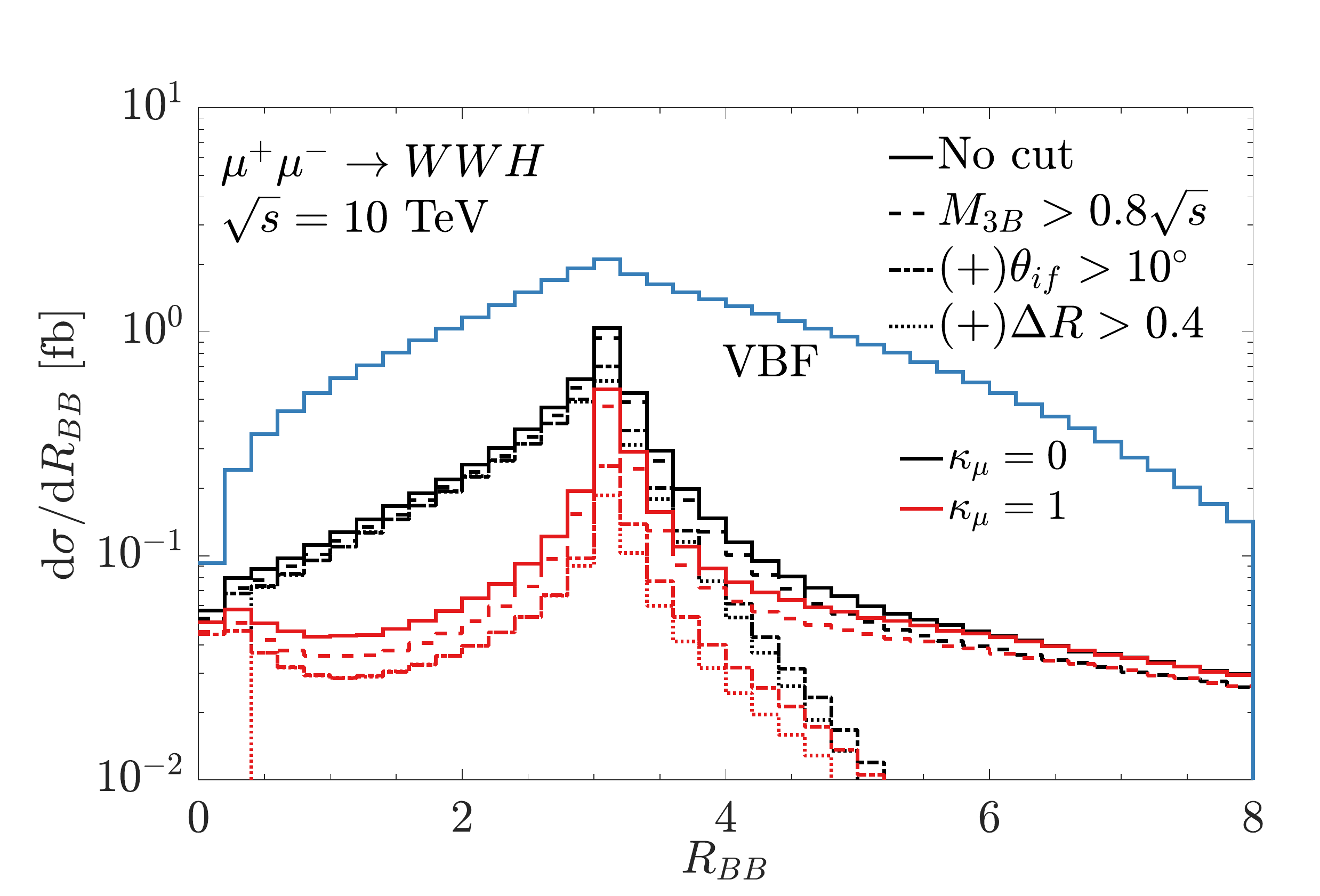}
         \includegraphics[width=0.48\textwidth]{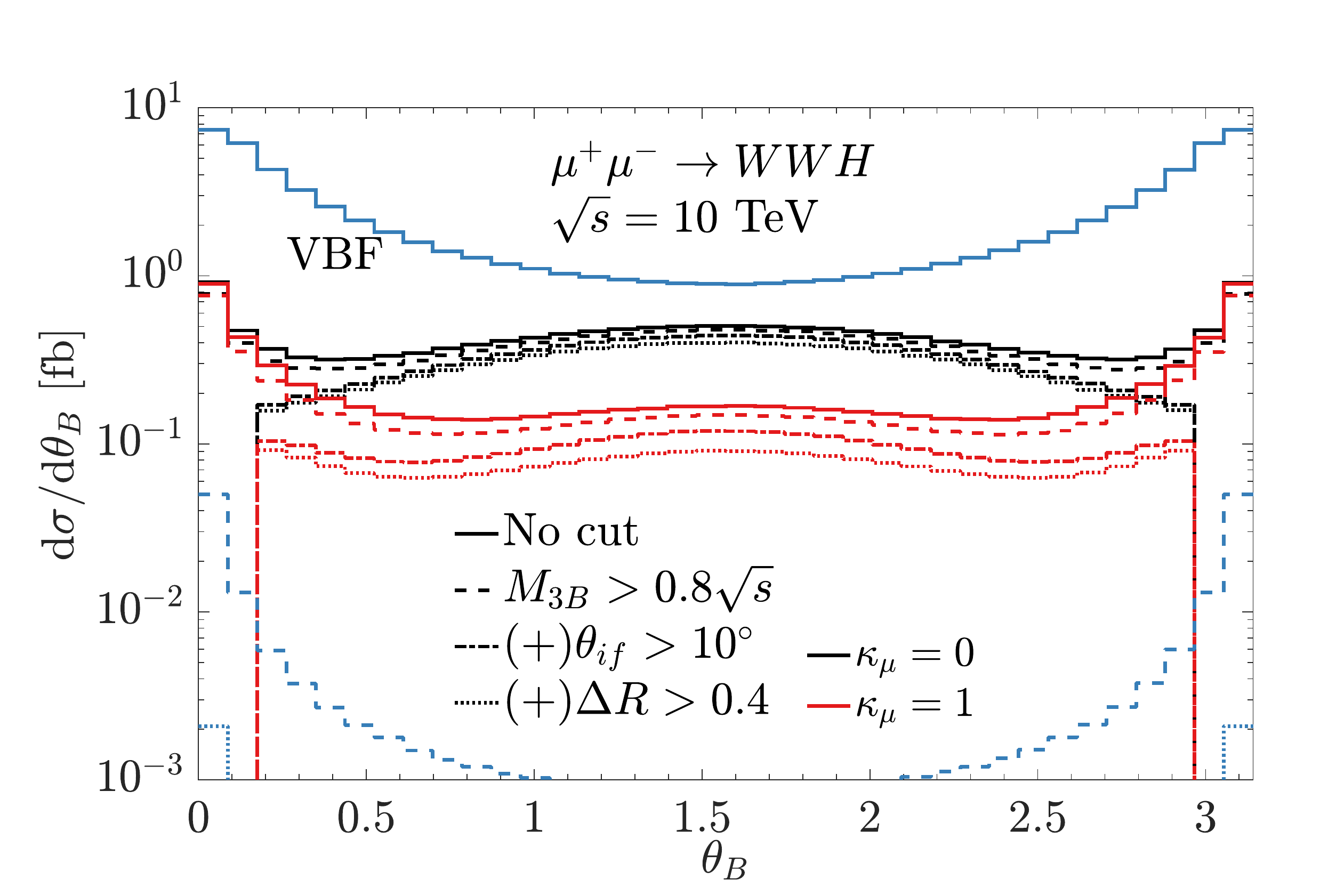}
         \caption{Differential distributions to isolate the prompt production from the VBF background.
         3-boson invariant mass (upper left), diboson $R$-distance (upper right), and boson polar angle (lower plot), respectively.}
         \label{fig:distributions}
\end{figure}

Besides their sensitivity on deviations in the gauge sector, the production of multiple Higgs and gauge bosons is also a very sensitive probe for the muon-Higgs coupling. In the SM, there is a subtle cancellation between muon mass insertion in the couplings to Higgs and longitudinal gauge bosons. In Fig.~\ref{fig:xsec}, we show the cross sections for the prompt production of one, two and three bosons (full lines), dashed lines show the cross sections of the corresponding final states in vector-boson fusion (VBF), and dotted lines show deviations due to the presence of higher-dimensional operators. In general, the deviations become the larger the higher the final-state multiplicity. However, in the same way the base cross sections become smaller, and the optimal sensitivity to deviations is given by the production of triple bosons, like $\mu^+\mu^- \to W^+W^-H , ZZH$ etc.  Triple-boson processes have been used in the same context at high-energy $e^+e^-$ colliders as well~\cite{Beyer:2006hx,Fleper:2016frz,BuarqueFranzosi:2021wrv}.
We have validated our framework by two independent analytic calculations using the Goldstone-boson equivalence theorem, and two different Monte Carlo generators, while the final cross sections and distributions for signal and background have been calculated with the event generator {\sc Whizard}~\cite{Kilian:2007gr}, using its FeynRules/UFO interface~\cite{Christensen:2008py} (note that recently it is also possible to generate these processes at next-to-leading (NLO) EW order~\cite{Bredt:2022dmm}) .

\begin{figure}[t]
\centering
    \includegraphics[width=.48\textwidth]{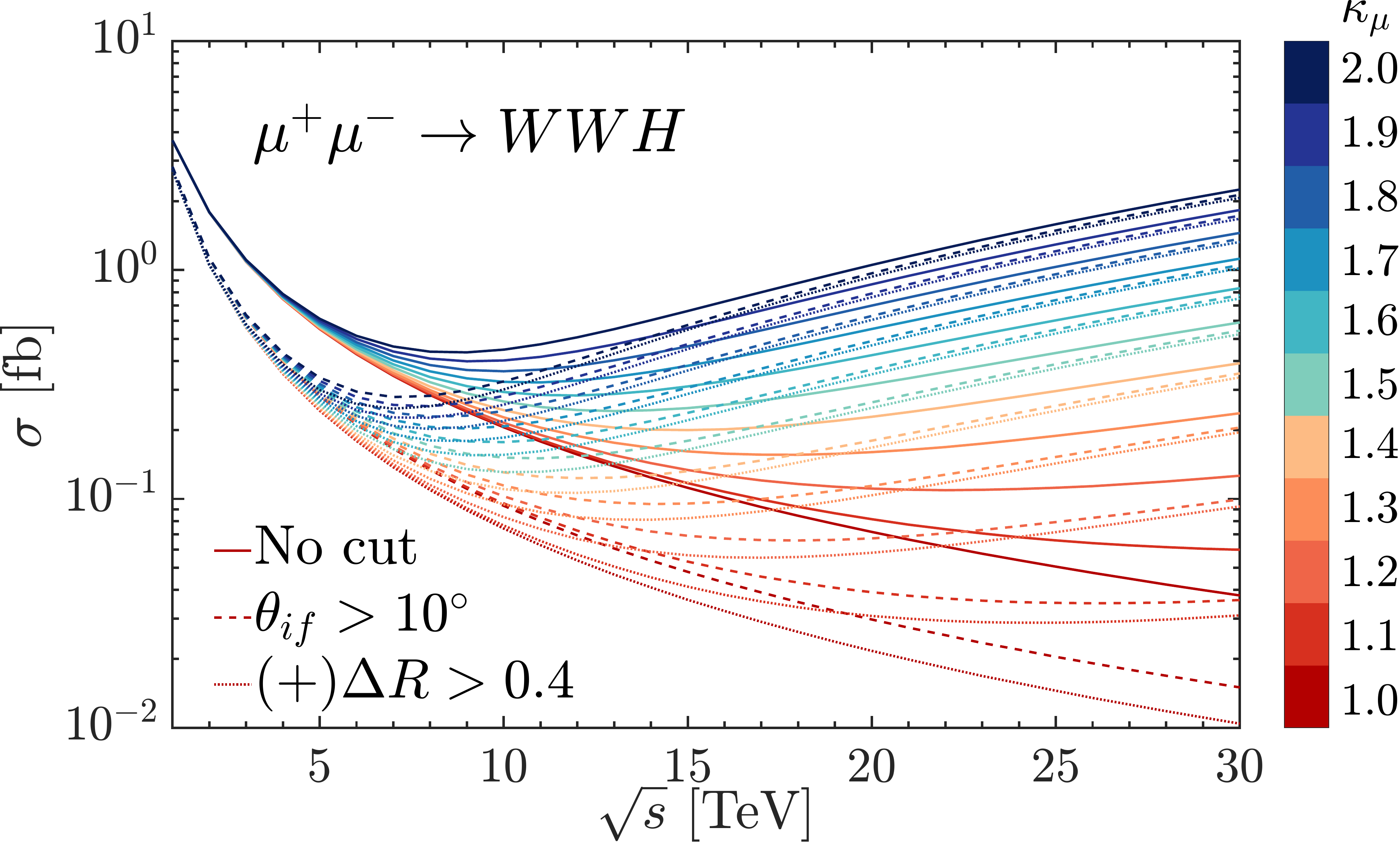}
    \quad
    \includegraphics[width=.48\textwidth]{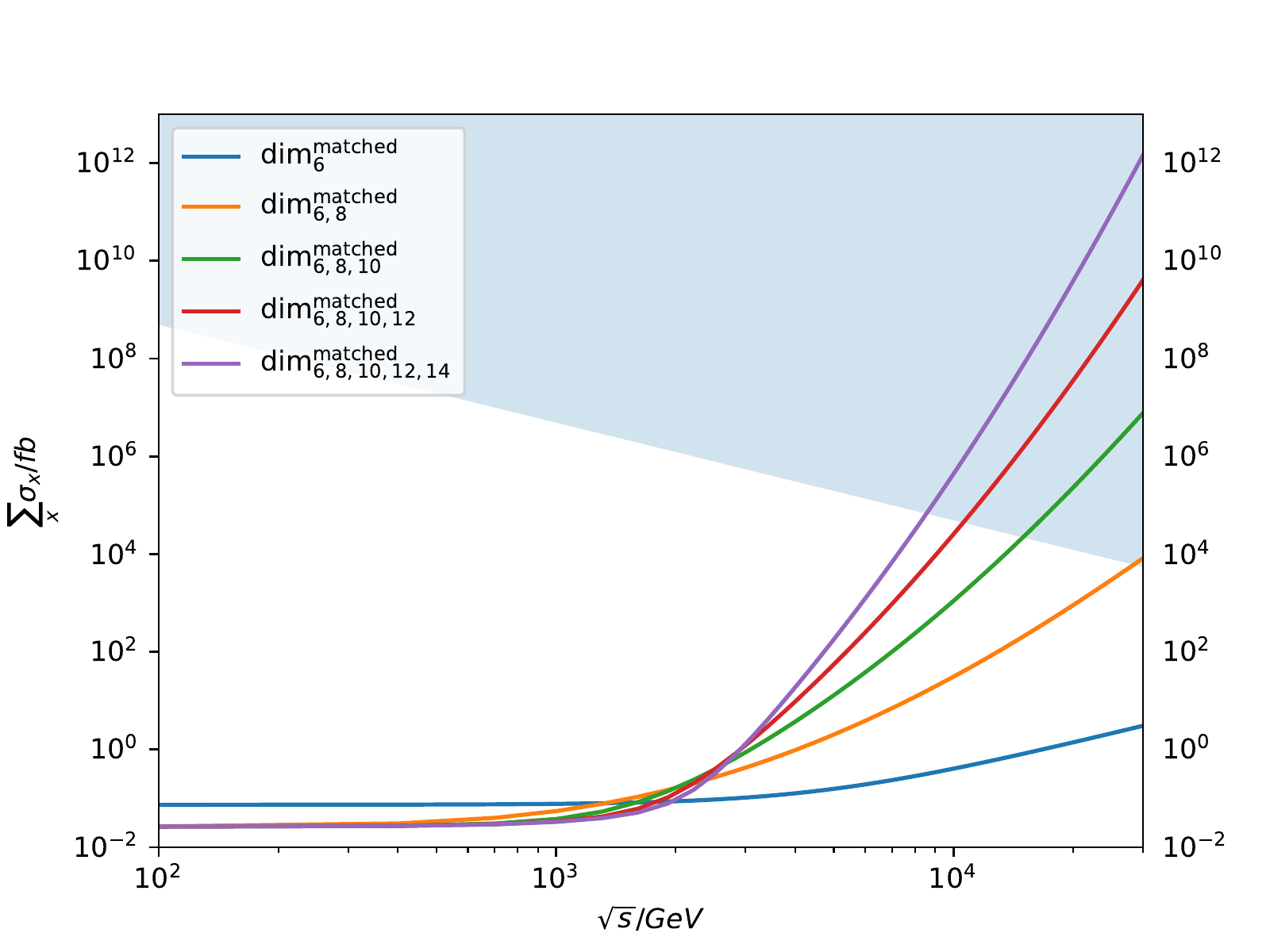}
    \caption{Cross section for the process $\mu^+\mu^- \to W^+W^-H$ as a function of the muon collider energy, with a variation of the muon-Higgs signal strength signifier, $\kappa_\mu$ (left); unitarity bound for the production of three Higgs and/or Goldstone bosons for different SMEFT truncation scenarios, in which the maximally possible multiplicity in the final state is 3,5,7,9,11 from bottom to top, respectively.}
    \label{fig:wwh_uni}
\end{figure}
At a high-energy muon collider, the EW boson fluxes from collinear splittings are high and hence there is a large rate of vector-boson
fusion/scattering (VBF/VBS) processes~\cite{Han:2020uid,Han:2021kes,Costantini:2020stv}. These, however, are not sensitive to the muon-Higgs
coupling like the prompt production, hence, the VBF processes must be kinematically suppressed with a dedicated cut flow. This is
demonstrated in Fig.~\ref{fig:distributions}: first of all, the triboson invariant mass which can be experimentally reconstructed very
well at a lepton collider, is peaked at the collider energy up to a tail from QED initial-state resummed photon radiation, while VBF
production peaks at threshold. This motivates a selection cut $M_{3B}>0.8\sqrt{s}$ to separate the two. Secondly, as VBF processes are
much more forward due to the collinear splitting, we employ a cut on the boson angles, $10\degree < \theta_B < 170\degree $ (also motivated by detector design).
Finally, to
separate the bosons, we apply a selection cut, demanding $\Delta R_{BB} > 0.4$. The effect of these cuts on the VBF and prompt
production is shown in Fig.~\ref{fig:distributions} for $M_{3B}$, $R_{BB}$ and $\theta_B$, respectively.

\begin{figure}[b]
\centering
    \includegraphics[width=0.75\textwidth]{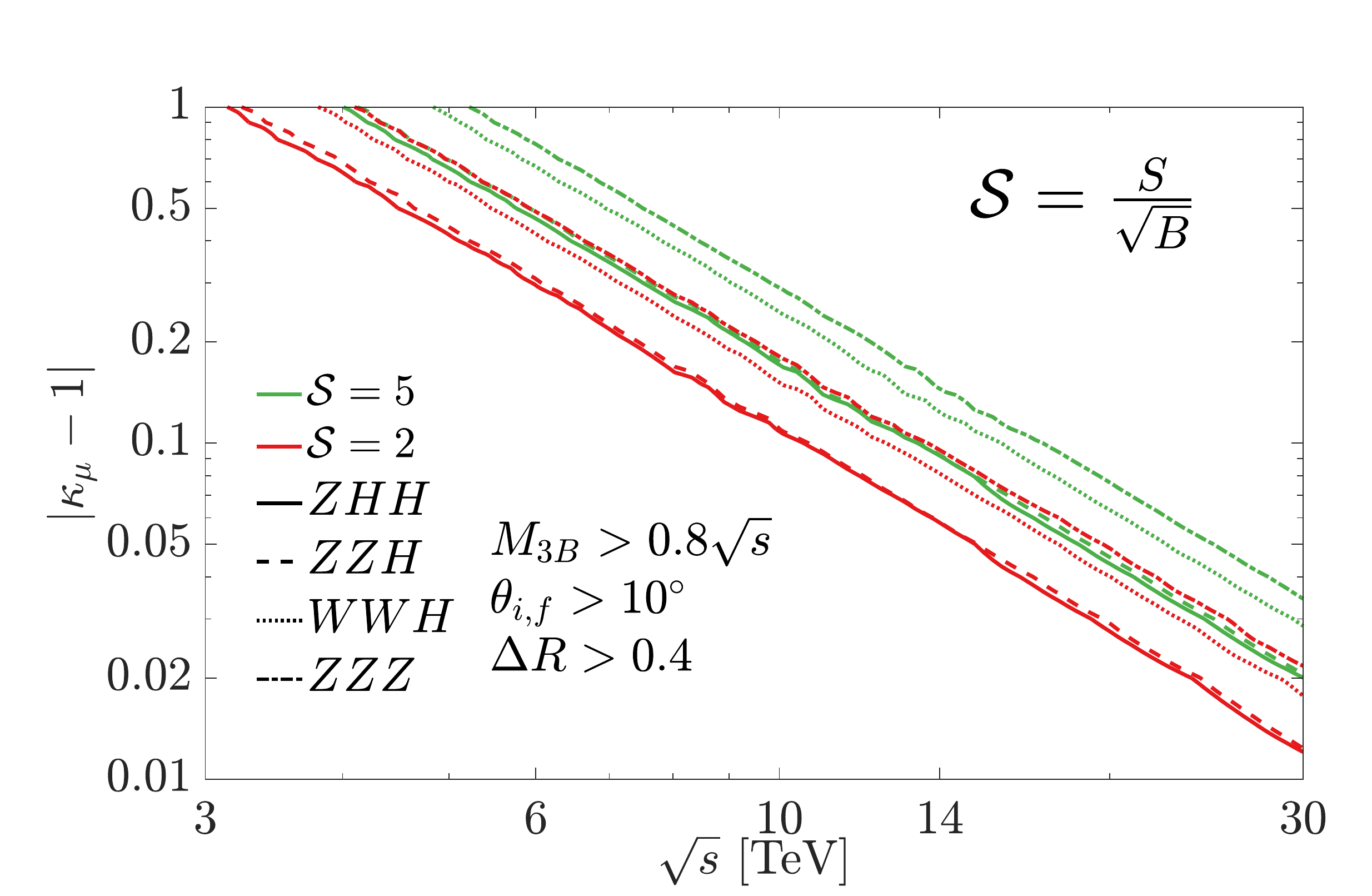}
    \caption{Statistical sensitivity to the muon-Higgs coupling $\kappa_\mu$ from three-boson measurements the muon collider.}
    \label{fig:significance}
\end{figure}
We concentrate here on the process $\mu^+\mu^- \to W^+W^-H$ which has not the largest, but a high sensitivity to deviations, and a large
(SM) cross section among the processes of multiplicity three. Fig.~\ref{fig:wwh_uni} shows on the left how the cross
section as a function of the collider energy varies with the muon-Higgs coupling modifier $\kappa_\mu$, which gives the ratio
$\kappa_\mu = Y_\mu / Y^{SM}_\mu$. On the right, the unitarity bound for the production of three Goldstone and/or Higgs bosons is shown
for different EFT scenarios described in detail in~\cite{Han:2021lnp}, based on the derivation in~\cite{Maltoni:2001dc}. This shows that for a 10 TeV muon collider there is no unitarity constraint, and for 30 TeV only in very extreme EFT scenarios.

As benchmark scenarios, we consider muon colliders in the energy range $1\,\text{TeV}< \sqrt{s} < 30\,\text{TeV}$ with the luminosity scaling $\mathcal{L} = \left(\sqrt{s} / 10\,\text{TeV} \right)^2 \cdot 10\,\text{ab}^{-1}$. We define the number of signal events as $S= N_{\kappa_\mu} - N_{\kappa_\mu = 1}$ and the number of background events $B = N_{\kappa_\mu = 1} + N_{\text{VBF}}$, respectively. Note that $N_{\kappa_\mu} \geq N_{\kappa_\mu  = 1}$. The statistical significance for anomalous muon-Higgs couplings is then defined as $\mathcal{S} = S/\sqrt{B}$. In Fig.~\ref{fig:significance} we show the statistical significance for a $2\sigma$ evidence and a $5\sigma$ discovery for the deviation from the SM muon-Higgs coupling, given by the deviation $\kappa_\mu - 1$ of the signal strength modifier from one. We show the significance from four different processes, $\mu^+\mu^- \to W^+W^-H, ZZH, ZHH, ZZZ$, respectively, using the cuts mentioned above to separate the signal from the VBF background. We can see that a 10 TeV collider is sensitive to a $\Delta\kappa_\mu/\kappa_\mu \sim 20\%$ deviation, while a 30 TeV collider could go down to 1-2\% deviation. Via the relation $\Lambda > 10\,\text{TeV} \cdot\sqrt{ g/ \Delta\kappa_\mu}$, where $g$ is the dimensionless Wilson coefficient in EFT framework, this translates into a discovery potential for new physics in the range of tens of TeV.
These results are to be put into relation with the sensitivity from the HL-LHC (and FCC-hh) which comes from the $H\to\mu\mu$ decay only which cannot resolve a case of $\kappa_\mu \sim - 1$.
\section{Conclusions}

In conclusion, a high-energy muon collider of 10-30 TeV, besides a plethora of fantastic physics capabilities for SM measurements and BSM discoveries, can access deviations of the muon-Higgs coupling from its SM value up to a precision of ten to a few percent, in a model-independent way. It thereby is able to probe physics scales of $\Lambda \sim 30 - 100$ TeV.

\section*{Acknowledgements}

JRR acknowledges the support by the Deutsche
Forschungsgemeinschaft (DFG, German Research Association) under
Germany's Excellence Strategy-EXC 2121 "Quantum Universe"-3908333.
This work has been supported in part by the U.S.~Department of Energy under grant No.~DE-FG02-95ER40896, U.S.~National Science Foundation under Grant No.~PHY-1820760, and in part by the PITT PACC.
This work has also been funded by the Deutsche Forschungsgemeinschaft (DFG,
German Research Foundation) - 491245950. WK and NK were supported in part by the Deutsche Forschungsgemeinschaft (DFG, German Research Foundation) under grant
396021762 – TRR 257.

\bibliographystyle{JHEP}
\bibliography{ichep_muon_yuk.bib}

\end{document}